\documentstyle[11pt,epsf,psfig,paspconf]{article}

\begin{document}

\title{Far--infrared Point Sources}
\author{B. Guiderdoni}
\affil{Institut d'Astrophysique de Paris, CNRS, 98bis Bld Arago, F--75014 
Paris}

\begin{abstract}
The analysis of the submm anisotropies that will be mapped by the 
forthcoming {\sc map} and {\sc planck} satellites requires careful 
foreground subtraction
before measuring CMB fluctuations. 
Among these, the foreground due to IR/submm thermal radiation from dusty
sources 
was poorly known until recent observational breakthroughs began unveiling the
properties of these objects. We hereafter briefly review the observational 
evidence for a strong evolution of IR/submm sources with respect to the 
local universe explored by {\sc iras}. We present the basic principles 
of a new 
modeling effort where consistent spectral energy distributions of galaxies 
are implemented into the paradigm of hierarchical clustering with the 
fashionable semi--analytic approach. This model provides us with specific 
predictions 
in IR/submm wavebands, that seem to reproduce the current 
status of the observations and help assessing the capabilities of forthcoming
instruments to pursue the exploration of the deep universe at IR/submm 
wavelengths. Finally, the ability of the {\sc planck} {\it High Frequency 
Instrument}
all--sky survey to produce a catalogue of dusty sources at submm wavelengths
is briefly described.
\end{abstract}


\section{Introduction}

The accurate measurement of the fluctuations of the Cosmic Microwave 
Background down to scales of a few arcmins by
the forthcoming satellite missions {\sc map} and {\sc planck} will 
require a thorough analysis of all the astrophysical sources of submm/mm 
anisotropies, and a careful separation of the various foreground 
components (Bouchet {\it et al.} 1996, Gispert \& Bouchet 1997,
Tegmark \& Efstathiou, 1996, 
Tegmark, 1998, Hobson {\it et al.} 1998a, Bouchet \& Gispert 1999). 
Among them, the foreground due to resolved/unresolved 
IR/submm galaxies that are present at all redshifts on the line of sight
was poorly known, until observations and analyses in the last three years
began unveiling the ``optically dark'' (and infrared--bright) side of galaxy 
evolution at cosmological distances. In parallel
to this observational breakthrough, a strong theoretical effort has opened up
the way to a modelling of these galaxies that was able to implement the
basic astrophysical processes ruling IR/submm emission
in a consistent way. As a consequence of this,
it is now possible to have a more general view on the problems of foreground 
separation, and on the capabilities of CMB missions to put new constraints
on the number and properties of these sources.

\begin{figure}[htb]
\centerline{
\psfig{figure=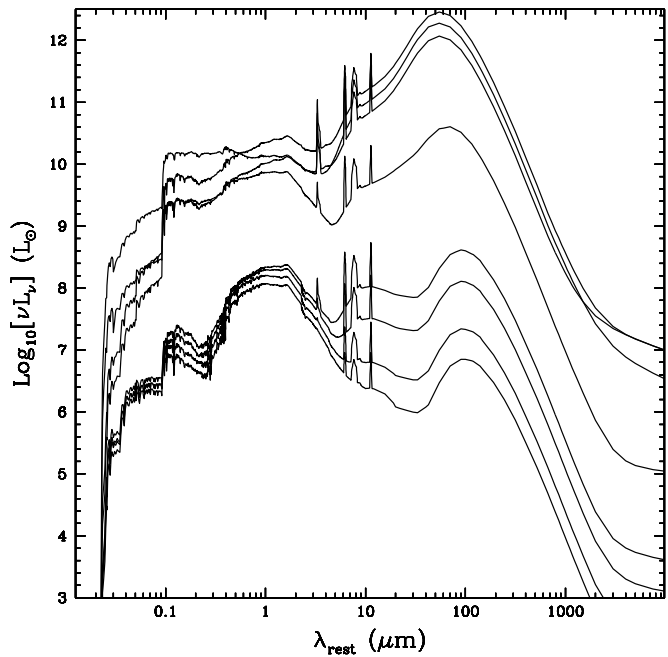,width=0.7\textwidth}}
\caption{\small A typical luminosity sequence for nearby spirals, 
LIRGs and ULIRGs,
in the spirit of fig. 2 in Sanders \& Mirabel (1996). The figure is taken from 
Devriendt {\it et al.} (1999). The IR luminosities range from $2.1 \times
10^7 L_\odot$ for the faintest spiral of the sample, to 
$6\times 10^{10} L_\odot $ for M82 (fifth curve from the bottom), 
and $4\times 10^{12}  L_\odot$ for the brightest ULIRG
of the sample.}
\end{figure}

In the {\it local} universe, we know from {\sc iras} observations that
about 30 \% of the bolometric luminosity of galaxies is 
radiated in the IR (Soifer \& Neugebauer 1991). Local galaxies can
be classified in a luminosity sequence from spirals (e.g the Milky 
Way -- the brightest spirals in the IR have a bar), and mild starbursts 
(e.g. M82), to the ``Luminous Infrared Galaxies'' (say,
with $10^{11} L_\odot < L_{IR} < 10^{12} L_\odot$),
and ``Ultra--Luminous Infrared Galaxies'' 
(say, with $10^{12}  L_\odot < L_{IR} $)
that radiate more than 95 \% of 
their bolometric luminosity in the IR/submm.

The IR/submm emission of these
sources is due to dust that absorbs UV and optical light, and thermally 
reradiates with a broad spectral energy distribution ranging from a few 
$\mu$m to a few mm. Most of the heating is due to young stellar populations
but, in the faintest objects, the average radiation field due to old 
stellar populations can be the main contributor, and, in the brightest 
objects (especially the ULIRGs), the question of the fraction of the heating 
that is due to a possible Active Galactic Nucleus is still difficult to 
assess. However, recent 
work based on {\sc iso} observations shows that starbursting still dominates
in 80 \% of ULIRGs, whereas the AGNs power only the 
brightest objects (Genzel {\it et al.} 1998, Lutz {\it et al.} 1998). 

Now, IR/submm observations are beginning to unveil what actually 
happened at higher redshift.
The detection of the ``Cosmic Infrared Background'' (hereafter CIRB)
at a level twice as high as 
the ``Cosmic Optical Background'' (hereafter COB) has shown that about 2/3 of
the luminosity budget of galaxies is emitted in the IR/submm range
(Puget {\it et al.} 1996).
In the same time, the first deep surveys at submm wavelengths have discovered 
the sources that are responsible for the CIRB, with a number density much
larger than the usual predictions based on our knowledge of the local 
universe (Smail {\it et al.} 1997). The optical follow--up of these sources 
is still in progress, 
but it appears that some (most ?) of them should be the moderate-- and 
high--redshift counterparts of the local LIRGs and ULIRGs
discovered by the {\sc iras} satellite and thoroughly studied by the 
{\sc iso} satellite. 

We shall hereafter focus on these dusty sources with thermal radiation,
and we refer the reader to the paper by Toffolatti and coworkers
(this volume), for the 
description of the foreground due to radiosources that emit free--free and 
synchrotron radiations at larger wavelengths. 

\begin{figure}[htb]
\centerline{ 
\psfig{figure=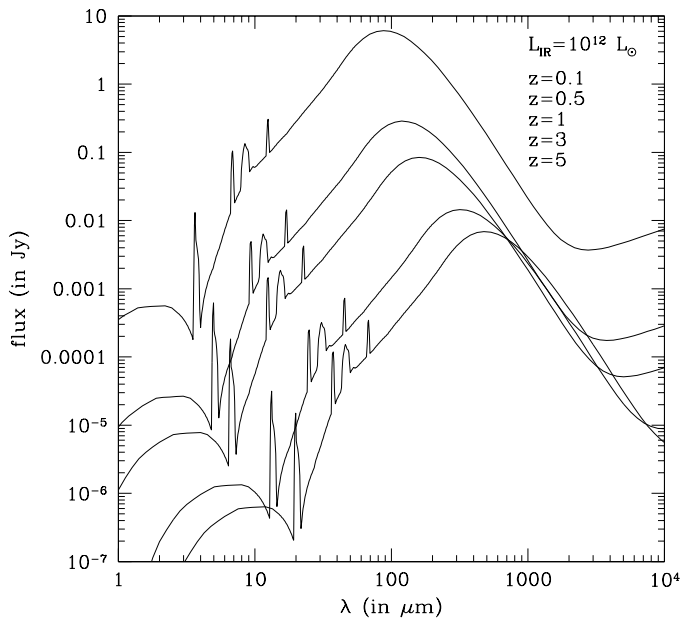,width=0.7\textwidth}}
\caption{\small Observer--frame model spectra of a $L_{IR}=10^{12} L_{\odot}$ 
galaxy at increasing redshifts (from top to bottom), for a cosmology
with $H_0=50$ km s$^{-1}$ Mpc$^{-1}$ and $\Omega_0=1$.
The reader is invited to note that the apparent flux
in the submm range is almost insensitive to redshift, because the shift of the
100 $\mu$m bump counterbalances distance dimming.}
\end{figure}

A consistent approach to the early evolution of galaxies is particularly
important for any attempt at predicting their submm properties. Three
basic problems have to be kept in mind, that explain why it is so difficult,
starting from general ideas about galaxy evolution, to get a correct 
assessment of 
the number density of faint submm sources and of the level of submm 
fluctuations they generate. 

First, it is difficult to extrapolate the IR/submm properties of galaxies 
from our knowledge of their optical properties. It is well known 
that there is no correlation between the optical and IR fluxes
-- see e.g. Soifer {\it et al.} (1987) for an analysis of the statistical
properties of the ``Bright Galaxy Sample''. 
Interestingly, the galaxies with the highest luminosities also emit most of 
their bolometric luminosity in the IR. 
If young stars are the dominant source of heating, it turns 
out that the strongest starbursts mainly emit in the IR/submm. Fig. 1
shows a sequence of spectral energy distributions for local galaxies with 
various IR luminosities, very much in the spirit of fig. 2 of Sanders \& 
Mirabel (1996). The sources have been modelled by Devriendt {\it et al.}
(1999, see section 5) under the assumption that starbursts are the dominant 
source of heating. Now, it is known that local LIRGs and ULIRGs
are interacting systems and mergers (e.g. Sanders \& Mirabel 1996). 
It is consequently plausible that
their number density should increase with redshift, when more fuel was 
available for star formation and more interactions could trigger it.
As a matter of fact, the Hubble Deep Field (HDF, Williams {\it et al.} 1996) 
has unveiled a large number of irregular/peculiar objects undergoing 
gravitational interactions (Abraham {\it et al.} 1996).
Such a large number of interacting systems is of course predicted by the
paradigm of hierarchical clustering, but the quantitative modelling of the 
merging rates of {\it galaxies}, and of the influence of merging
on star formation is highly uncertain.
\begin{figure}[htb]
\centerline{
\psfig{figure=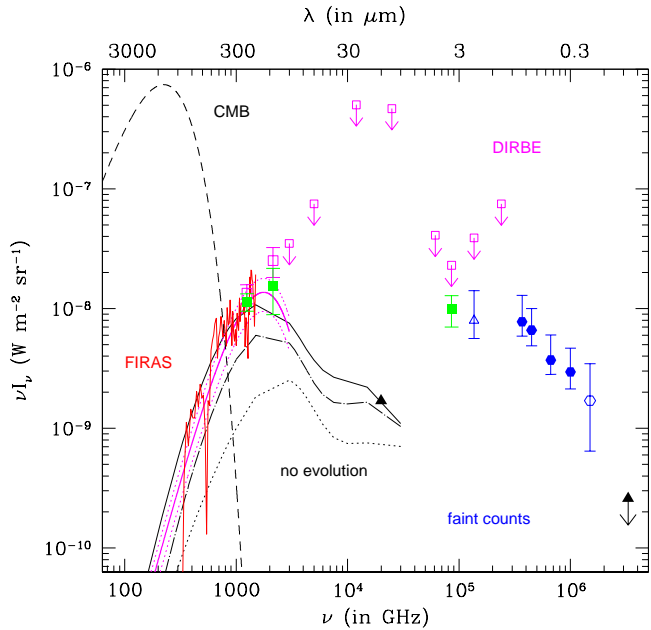,width=0.7\textwidth}}
\caption{\small The Cosmic Optical and Infrared Backgrounds. 
Open triangle, solid dots, and open dot : COB obtained from faint 
counts and compiled by Pozzetti {\it et al.} (1998). Solid triangle at 
0.0912 $\mu$m : upper limit by Vogel {\it et al.} (1995). 
Solid triangle at 15 $\mu$m : Oliver {\it et al.} (1997). 
Thick solid line : CIRB extracted by Fixsen {\it et al.} (1998). 
Thin dotted lines : error bars. Broken solid line : CIRB in the
cleanest regions of the sky (Puget {\it et al.} 1996, Guiderdoni {\it et al.} 
1997). Open squares : upper limits and detections (at 140 and 240 $\mu$m) 
from Hauser {\it et al.} (1998). Solid squares : reanalysis at 100 and 240 
$\mu$m (Lagache {\it et al.} 1999) and 3.5 $\mu$m (Dwek \& Arendt 1998).
The no--evolution curve (dotted line) is computed from the 
local {\sc iras} luminosity function extrapolated to $z=8$, 
for $H_0=50$ km s$^{-1}$ 
Mpc$^{-1}$ and $\Omega_0=1$.  The thin solid line shows the best model 
(the so--called ``model E'') in Guiderdoni {\it et al.} (1998). Dashes and
dots give ``model A''.}
\end{figure}

Second, we might have kept so far the prejudice that high--redshift
galaxies have little extinction, simply because their heavy--element 
abundances are low (typically 1/100 to 1/10 of solar at $z>2$). However, low 
abundances do not necessarily mean low extinction. For instance, if we 
assume that dust grains have a size distribution similar to the one of our 
Galaxy 
($n(a)da \propto a^{-3.5}$ with $a_{min} \le a \le a_{max}$), and are
homogeneously distributed in a region with radius $R$, the optical depth varies
as $\tau \propto a_{min}^{-0.5}R$, whereas the total dust mass varies as  
$M_{dust} \propto a_{max}^{0.5}R^3$. For a given dust mass and size 
distribution, there is more extinction where grains are small, and 
close to the heating sources. This is probably the reason why
Thuan {\it et al.} (1998) observed a 
significant dust emission in the extremely metal--poor galaxy SBS0335-052.
In this context, modelling chemical evolution and transfer is not an easy task.

Third, distant galaxies are readily observable at submm wavelengths.
Fig. 2 shows {\it model} spectra of an ULIRG as it would be 
observed if placed at different
redshifts. There is a wavelength range, between $\sim 600$ $\mu$m and $\sim 4$
mm, in which the distance effect is counterbalanced by the ``negative
k--correction'' due to the rest--frame emission maximum at $\sim 100$
$\mu$m. In this range, the apparent flux of galaxies depends weakly on
redshift to the point that, evolution aside, a galaxy might be easier to
detect at $z=5$ than at $z=0.5$. The observer--frame submm fluxes, faint
galaxy counts, and diffuse background of unresolved galaxies are consequently
very sensitive to the early stages of galaxy evolution. Note that this
particular wavelength range brackets the maximum of emission of the CMB.
As a consequence, any uncertainty in the modelling of galaxy evolution at 
high $z$ will strongly reflect on the results of the faint counts of resolved
sources, and on the fluctuations of the foreground of unresolved sources.

In sections 2 and 3, we report respectively on the recent observation of the 
CIRB, and the faint submm counts with the {\sc iso} satellite and the 
{\it SCUBA}
instrument on the James Clerk Maxwell Telescope (see the review by Mann 
and coworkers in 
this volume). In section 4, we briefly mention the efforts to correct the 
optical surveys for the effect of extinction, that give a lower limit of 
the number of submm sources from the number of sources detected at optical 
wavelengths. In section 5, we resume various attempts developped so far 
to compute consistent optical/IR spectra, and to model  
IR/submm counts. In section 6, we summarize recent developments of the 
semi--analytic modelling of galaxy formation and evolution where the 
computation of dust extinction and emission is explicitly implemented.
Finally, in section 7, we sketch an overview of the sensitivities of 
forthcoming instruments that should greatly improve our knowledge of 
IR/submm sources, and we emphasize the capability of the {\sc planck} 
{\it High Frequency Instrument} to get an all--sky survey of bright, 
dusty sources at submm wavelengths.

\begin{figure}[htb]
\vspace{0.5truecm}
\centerline{
\psfig{figure=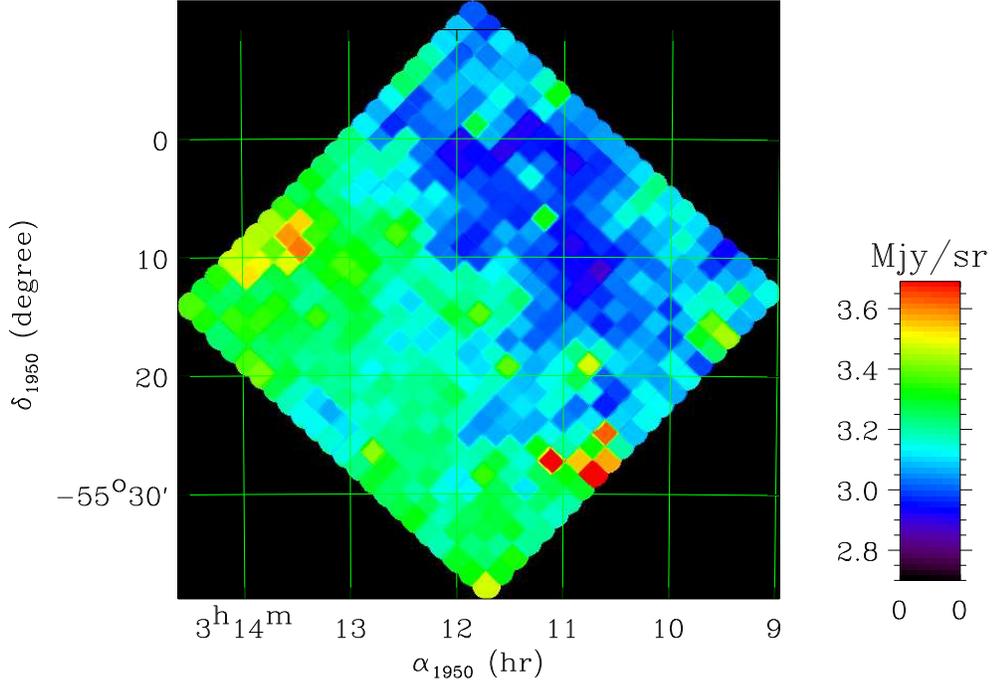,width=\textwidth}}
\caption{\small One of the Southern fields observed by {\it ISOPHOT}
at 175 $\mu$m for the FIRBACK program (from Puget {\it et al.} 1999). 
The pixel size is 1.5 arcmin. The large--scale fluctuations are due 
to the presence of Galactic cirrus, and 24 sources have been identified at 
$S_{175} > 100$ mJy.}
\end{figure}

\section{The Cosmic Infrared Background}
The epoch of galaxy formation can be observed by its imprint on
the background radiation that is produced by the accumulation
of the light of extragalactic sources along the line of sight. 
The direct search for the COB currently gives only upper limits. 
However, estimates of lower limits can be obtained by summing up the 
contributions of faint galaxies. The shallowing
of the faint counts obtained in the HDF (Williams {\it et al.} 1996) 
suggests that these lower limits are close to convergence. 

In the submm range, Puget {\it et al.} (1996) have
discovered an isotropic component in the {\it FIRAS} residuals
between 200 $\mu$m and 2 mm. This measure was confirmed 
by subsequent work in the cleanest regions of the sky (Guiderdoni {\it et al.}
1997), and by an independent determination (Fixsen {\it et al.} 1998), giving
a mean value of the background 
$\nu I_\nu = 1.3 \times 10^{-5} (\lambda_{100})^{-0.64}
\nu B_\nu(T_d=18.5{\rm K})$ where $\lambda_{100}$ is the wavelength in units 
of 100 $\mu$m. The analysis of the {\it DIRBE}
dark sky has also led to the detection of the isotropic background at 240 and 
140 $\mu$m, and to upper limits at shorter wavelengths down to 2 $\mu$m
(Schlegel {\it et al.} 1998, Hauser {\it et al.} 1998). Recently, a measure
at 3.5 $\mu$m was proposed by Dwek \& Arendt (1998).
The results of these analyses seem in good agreement, though the exact level 
of the background  around 140 and 240 $\mu$m is still a matter of debate. The 
controversy concerns the correction for the amount of Galactic dust in
the ionized gas uncorrelated with the HI gas. A 
new assessment of the issue by Lagache {\it et al.} (1999) leads to values
of the CIRB that are in good agreement with the fit of {\it FIRAS} data by 
Fixsen {\it et al.} (1998), and to values at 140 and 240 $\mu$m that are
lower than in Hauser {\it et al.} (1998).
Figure 3 displays the various determinations.

\begin{figure}[htb]
\centerline{
\psfig{figure=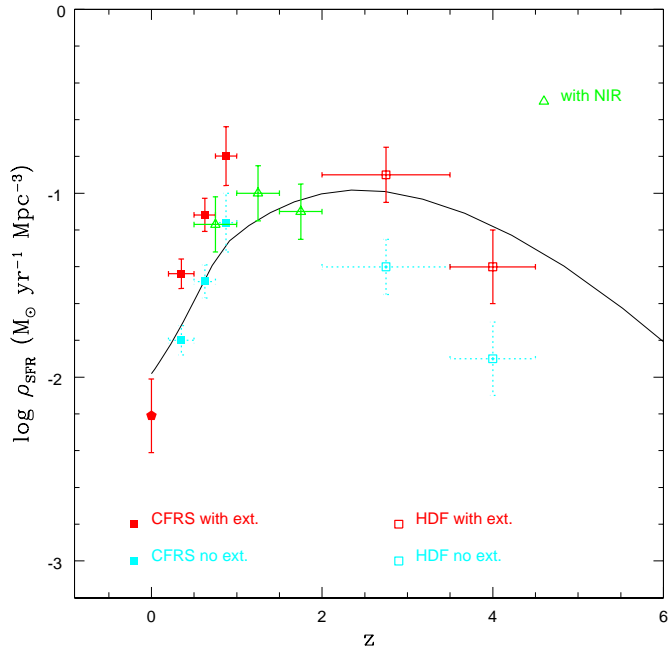,width=0.7\textwidth}}
\caption{\small The evolution of the 
cosmic Star Formation Rate comoving density $\rho_{SFR}$ with redshift $z$. 
Solid pentagon : local value (Gallego {\it et al.} 1995).
Solid dots and error bars (dotted lines) : CFRS uncorrected for extinction 
(Lilly {\it et al.} 1996). Solid dots and error bars (solid lines) : 
multi--wavelength analysis including IR,
submm, and radio data (Flores {\it et al.} 1999). 
Open squares and error bars (dotted lines) : HDF uncorrected for extinction 
(Madau {\it et al.} 1996, 1998).
Open squares and error bars (solid lines) : values corrected 
according to Pettini {\it et al.} (1998).
The corrections derived by Meurer {\it et al.} (1997) would shift the 
corrected points upwards by $\sim$0.5 dex. Open triangles :
(Connolly {\it et al.} 1997). The rest--frame UV fluxes are converted into 
SFRs according to Guiderdoni {\it et al.} (1998), and a Salpeter IMF. 
Solid line : best model in the latter paper (the so--called ``model E'').}
\end{figure}

It appears very likely that this isotropic background is the long--sought
CIRB (Puget {\it et al.} 1996, 
Dwek {\it et al.} 1998). As shown in fig. 3, its level 
is about 5--10 times the no--evolution prediction based on the local IR
luminosity function determined by {\sc iras}. There is about twice as much flux
in the CIRB than in the COB. If the dust that emits at IR/submm wavelengths 
is mainly heated by young stellar populations, the sum of the fluxes of the 
CIRB and COB gives the level of the Cosmic Background associated with
stellar nucleosynthesis (Partridge \& Peebles 1967). The bolometric 
intensity (in W m$^{-2}$ sr$^{-1}$) is~:
\begin{equation}
I_{bol}=\int {\epsilon_{bol} \over 4\pi} {dl \over (1+z)^4} = {c\eta
\over 4\pi}{\rho_Z(z=0) \over (1+z_{eff})}
\end{equation}
where $\epsilon_{bol}(t)=\eta (1+z)^3 \dot \rho_Z(t)$ is the physical 
emissivity 
due to young stars at cosmic time $t$, and $z_{eff}$ is the effective 
redshift for stellar He and metal nucleosynthesis.
An approximate census of the local density of heavy elements $\rho_Z(z=0)
\sim 1 \times 10^7$ $M_\odot$ Mpc$^{-3}$, taking into account the metals in
the hot gas of rich galaxy clusters (Mushotzky \& Loewenstein 1997)
gives an expected bolometric
intensity of the background $I_{bol} \simeq 50(1+z_{eff})^{-1}$ 
nW m$^{-2}$ sr$^{-1}$. This value is roughly consistent with the observations 
for $z_{eff} \sim 1$ -- 2.
\begin{figure}[htb]
\centerline{ 
\psfig{figure=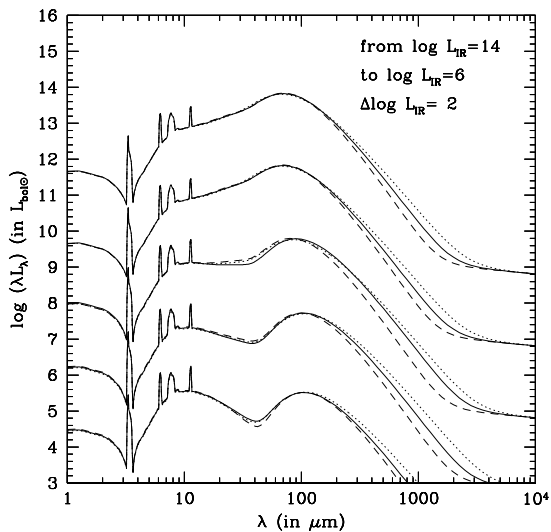,width=0.6\textwidth}}
\caption{\small Model spectra in the IR and submm, for IR luminosities $10^6$,
$10^8$, $10^{10}$, $10^{12}$, and $10^{14} L_\odot$.  Emissivity index of
big grains : $m=2$ (dashes), $m=1.5$ (solid lines), $m=1$ (dotted lines).
The sequence is drawn from Guiderdoni {\it et al.} (1998).}
\end{figure}

Of course, it is not clear yet whether star formation is responsible
for the bulk of dust heating, or there is a significant contribution of AGNs. 
In order to address this issue, one has first to identify 
the sources that are responsible for the CIRB.
The question of the origin of dust heating in heavily--extinguished 
objects is a difficult one, because both starburst and AGN rejuvenation 
can be fueled by gas inflows triggered by interaction, and IR/submm
spectra can be very similar if extinction is large.
However, according to Genzel {\it et al.} (1998), 
the starburst generally contributes to 50--90 \% of the heating in local 
ULIRGs. About 80 \% of the ULIRGs in the larger local sample of 
Lutz {\it et al.} 
(1998) are dominated by the starburst, but the trend decreases with 
increasing luminosity, and the brightest objects are AGN--dominated.
Now, it is very likely that the 
high--redshift counterparts of the local LIRGs and ULIRGs are responsible 
for the CIRB. However the redshift evolution of the fraction and power of AGNs 
that are harbored in these distant objects is still unknown.

\section{Far--infrared galaxies at high redshift}
Various submm surveys have been achieved or are in progress.
The FIRBACK program is a deep survey of 4 deg$^2$ at 175 $\mu$m with the 
{\it ISOPHOT} instrument aboard {\sc iso}. The analysis of about 1/4 of 
the Southern 
fields (that is, of 0.25 deg$^2$, see fig. 4) unveils 24 sources 
(with a $5\sigma$ flux limit $S_\nu > 100$ mJy), 
corresponding to a surface density five times larger 
than the no--evolution predictions based on the local IR luminosity 
function (Puget {\it et al.} 1999). It is likely that we are actualy
seeing the maximum emission bump at 50--100 $\mu$m redshifted at cosmological
distances, rather than a local population of sources with a very cold dust
component, which seems to be absent from the shallow {\it ISOPHOT} survey at 
175 $\mu$m (Stickel {\it et al.} 1998).
The total catalogue of the 4 deg$^2$ will
include about 275 sources (Dole {\it et al.} 1999). The radio
and optical follow--up for identification is still in progress. 
This strong evolution is confirmed
by the other 175 $\mu$m deep survey by Kawara {\it et al.} (1998).
 
The {\it ISOCAM} deep surveys at 15 $\mu$m also conclude to a significant 
evolution of the sources (Oliver {\it et al.} 1997, Aussel {\it et al.} 1998,
Elbaz {\it et al.} 1998, 1999). 
Most of the sources identified so far by the optical follow--up  
have typical redshifts $z \sim 0.7$, and optical 
colours similar to those of field galaxies, with morphologies that frequently
have signs of interaction. The surveys seem to show a population of bright
peculiar galaxies, starbursts, LIRGs, and AGNs.
The observer--frame 15 $\mu$m waveband 
corresponds to rest--frame wavelengths that probe the properties of PAH and 
very small grains, at the depth of the survey. The extent to which 
the 15 $\mu$m flux is related 
to the bulk of IR/submm emission produced by star formation is under study. 
 
Various deep surveys
at 850 $\mu$m have been achieved with the {\it SCUBA} instrument at the JCMT
(Smail {\it et al.} 1997, Hughes {\it et al.} 1998, Barger {\it et al.} 1998, 
Eales {\it et al.} 1998). They also unveil a surface density of sources
(with $S_\nu > 2$ mJy) much larger than the no--evolution predictions
(by two or three orders of magnitude !).
The total number of sources discovered in {\it SCUBA} deep surveys 
now reaches about 35 (see e.g. Blain {\it et al.} 1998) and should rapidly 
increase. The tentative optical identifications seem to show that 
these objects look like distant LIRGs and 
ULIRGs (Smail {\it et al.} 1998, Lilly {\it et al.} 1999). 
In the HDF, 4 of the brightest 5 sources seem to lie
between redshifts 2 and 4 (Hughes {\it et al.} 1998), but the 
optical identifications are still a matter of debate 
(Richards, 1998). The source SMM 02399-0136 at $z=2.803$, which is
gravitationally amplified by the foreground cluster A370, 
is clearly an AGN/starburst galaxy 
(Ivison {\it et al.} 1998, Frayer {\it et al.} 1998).

\section{The optical view and the issue of extinction}
Recent observational breakthroughs have made possible the measurement
of the Star Formation Rate (SFR) history of the universe from rest--frame 
UV fluxes of moderate-- and high--redshift galaxies
(Lilly {\it et al.} 1996,
Madau {\it et al.} 1996, 1998, Steidel \& Hamilton 1993, Steidel 
{\it et al.} 1996, 1999).
Since the early versions of the reconstruction of the cosmic SFR density, 
much work has been done to address dust issues. However, a complete assessment
of the effect of extinction on UV fluxes emitted by young
stellar populations, and of the luminosity budget of star--forming galaxies
is still to come. Dust seems to be present even at large redshifts, since the
optical spectrum of a gravitationally--lensed galaxy at $z=4.92$ 
(Franx {\it et al.} 1997) already shows a reddening factor amounting 
to $0.1 < E(B-V) < 0.3$ (Soifer {\it et al.} 1998).

\begin{figure}[htb]
\centerline{
\psfig{figure=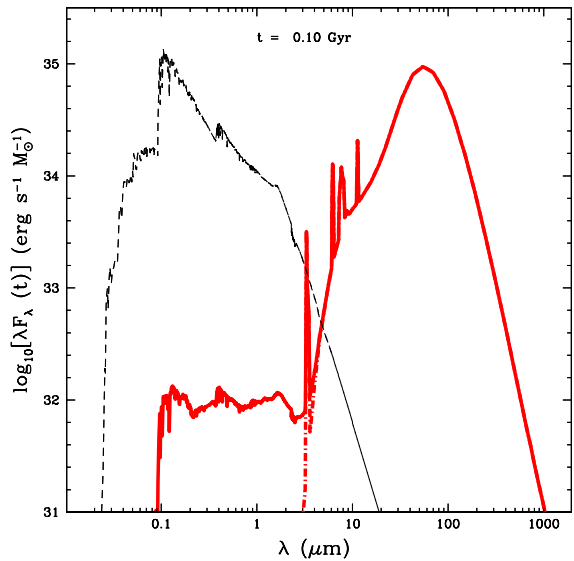,width=0.6\textwidth}}
\caption{\small Spectral energy distribution of an ULIRG, computed by
Devriendt {\it et al.} (1999). The thin solid line shows the typical 
spectrum of a starburst at age 0.10 Gyr with the same bolometric luminosity. 
The thick solid line and dashes give the spectrum
with two different geometries for the dust and star distribution : an
oblate spheroid (preferred geomety), and a screen model. Whereas there 
is almost no difference in 
the IR/submm part, the predicted optical properties are {\it very} 
sensitive to the dust distribution.}
\end{figure}

The cosmic SFR density determined only from the UV fluxes of the 
Canada--France Redshift Survey has been recently revisited 
with optical, IR, and radio observations. 
The result is an upward correction of the previous values by an 
average factor 2.9 (Flores {\it et al.} 1999). 
At higher redshift, various authors have attempted to estimate
the extinction correction and to recover the fraction of UV starlight 
absorbed by dust (e.g. Meurer {\it et al.} 1997, Pettini {\it et al.} 1998). 
It turns out 
that the observed slope $\alpha$ of the UV spectral energy distribution
$F_\lambda (\lambda) \propto \lambda^\alpha$ (say, around 2200 \AA)
is flatter than the standard value  $\alpha_0 \simeq -2.5$ 
computed from models of spectrophotometric evolution. The derived 
extinction 
corrections are large and differ according to the method. For instance,
Pettini {\it et al.} (1998) and coworkers fit a typical extinction curve 
(the Small Magellanic Cloud one) to the observed colors,
whereas Meurer {\it et al.} (1997) and coworkers use an 
empirical relation between $\alpha$ and the FIR to 2200 \AA\ luminosity ratio
in {\it local} starbursts. The former authors
derive $<E(B-V)> \simeq 0.09$ resulting 
in a factor 2.7 absorption at 1600 \AA, whereas the latter 
derive $<E(B-V)> \simeq 0.30$ resulting in a 
factor 10 absorption. This discrepancy suggests sort of
a bimodal distribution of the young stellar populations : the first 
method would take into account the stars detected in the UV with relatively
moderate reddening/extinction, while the 
second one would phenomenologically add the contributions of these ``apparent''
stars and of heavily--extinguished stars.
Fig. 5 shows the cosmic SFR comoving density in the early version
(no extinction), and after the work by Flores {\it et al.} (1999) 
at $z<1$ and the extinction correction derived by Pettini {\it et al.} (1998) 
at higher redshift. 

The broad maximum observed at $z \sim 1.5$ to 3 (see fig. 5) seems to 
be correlated 
with the decrease of the cold--gas comoving density in damped Lyman--$\alpha$ 
systems between $z=2$ and $z=0$ (Lanzetta {\it et al.} 1995,
Storrie--Lombardi {\it et al.} 1996). These results nicely fit in 
a view where star formation in bursts triggered by interaction/merging
consumes and enriches the gas content of galaxies as time goes on.
It is common wisdom that such a qualitative scenario is
expected within the paradigm of hierarchical growth of structures. The
implementation of hierarchical galaxy formation in semi--analytic models
confirms this expectation (e.g. Baugh {\it et al.} 1998, and references 
therein). The question is to 
know whether the observations of the optically dark side of galaxies 
could modify this view significantly.

\begin{figure}[htb]
\centerline{
\psfig{figure=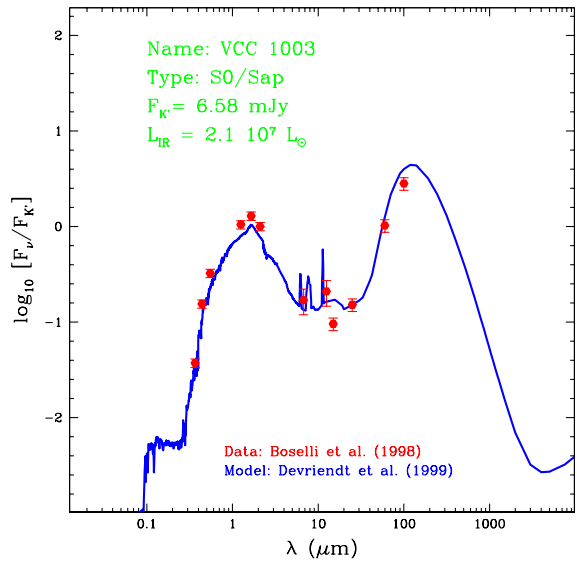,width=0.6\textwidth}}
\caption{\small Spectral energy distribution of a spiral galaxy. The solid
line shows the fit of the data with a theoretical spectrum computed by
Devriendt {\it et al.} (1999).}
\end{figure}

\section{Modelling dust spectra and IR/submm counts}
Various models have been proposed to account for the IR/submm emission of 
galaxies and to predict forthcoming observations. The level of 
sophistication (and complexity) increases from pure luminosity and/or density 
evolution extrapolated from the {\sc iras} local luminosity function 
with $(1+z)^n$ 
laws, and modified black--body spectra, to physically--motivated spectral 
evolution.

\begin{figure}[htb]
\centerline{
\psfig{figure=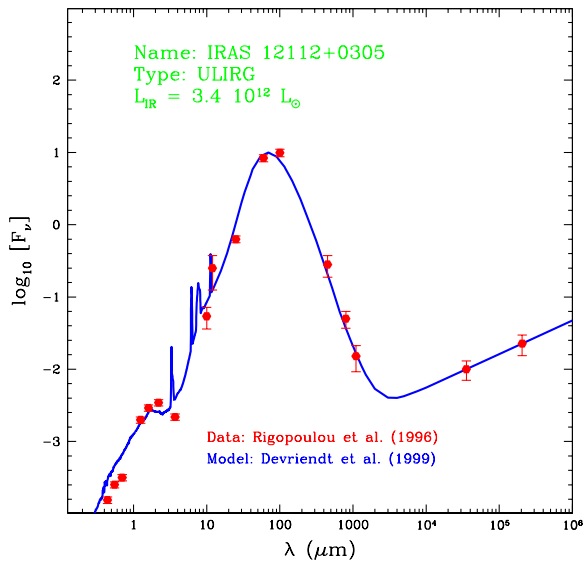,width=0.6\textwidth}}
\caption{\small Spectral energy distribution of an ULIRG. The solid
line shows the fit of the data with a theoretical spectrum computed by
Devriendt {\it et al.} (1999).}
\end{figure}

The evolution of the IR/submm luminosities can be
computed from the usual modelling of spectrophotometric evolution, by 
implementing the involved physical processes (stellar evolutionary
tracks and stellar spectra, chemical evolution, 
dust formation, dust heating and transfer, dust thermal emission).
Dwek (1998) tried to explicitly model the processes of dust formation and 
destruction (see references therein for a review of this complicated
issue). Most models prefer to assume simple relations between the dust content 
and the heavy--element abundance of the gas. The simplest assumption is a 
dust--to--gas ratio that is proportional to the heavy--element abundances.

\begin{figure}[htb]
\centerline{
\psfig{figure=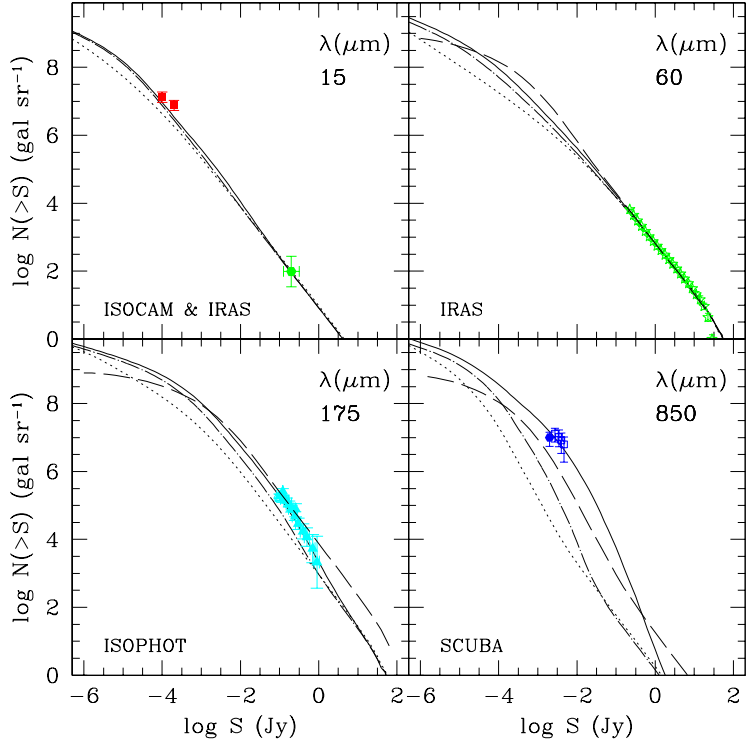,width=0.8\textwidth}}
\caption{\small Predicted counts at 15 $\mu$m, 60 $\mu$m, 175 $\mu$m, and 
850 $\mu$m for models A (dashes and dots) and E (solid line). Model E is 
normalized to 
reproduce the CIRB. Dotted line : no evolution. Dashes : FIR/submm thermal
sources in Toffolatti {\it et al.} (1998).
Solid squares : {\sc iso}--HDF at 15 $\mu$m (Oliver {\it et al.} 1997) ; 
solid hexagon : Rush {\it et al.} (1993).
Open stars : Faint Source Survey (Lonsdale {\it et al.} 1990).
Open hexagon : {\it ISOPHOT} at 175 $\mu$m (Kawara {\it et al.} 1998) ;
triangles : {\it ISOPHOT} at 175 $\mu$m (Puget {\it et al.} 1999, 
Dole {\it et al.}
1999). Open squares : deep {\it SCUBA} survey (Smail {\it et al.} 1997) ; 
solid dot : {\it SCUBA}--HDF (Hughes {\it et al.} 1998).}
\end{figure}

Guiderdoni {\it et al.} (1996, 1997, 1998) proposed a consistent modelling of 
IR/submm spectra that was designed to be subsequently implemented in
semi--analytic models of galaxy formation and evolution (see section 6).
The values of the free parameters that appear in this modelling (gas mass
and metallicity, 
radius of the gaseous disk) are readily computable in semi--analytic models
for the overall population of galaxies.
The IR/submm spectra of galaxies are computed according to 
Guiderdoni \& Rocca--Volmerange (1987), as follows : 
\begin{enumerate}
\item follow chemical evolution of the gas ;
\item implement extinction curves which depend 
on metallicity according to observations in the Milky Way, the LMC and SMC ;
\item compute $\tau_\lambda \propto Z_{gas}^s N_{gas} (A_\lambda/A_V)$
where $Z_{gas}$ and $N_{gas}$ are the gas metallicity and column density, 
$s=1.6$ for $\lambda >$ 2000 \AA\ (and 1.35 below), and $A_\lambda/A_V$
is the Milky Way extinction curve ;
\item assume the so--called ``slab'' or oblate spheroid geometries where 
the star and dust components are homogeneously mixed with equal height scales. 
The choice of these simple geometries for transfer is motivated by 
studies of nearby samples 
(Andreani \& Franceschini 1996) ;
\item compute a spectral energy distribution by assuming a mix of 
various dust components (PAH, very small grains, big grains) according
to D\'esert {\it et al.} (1990). The contributions are fixed in order 
to reproduce the observational correlation of {\sc iras} colours with total 
IR luminosity (Soifer \& Neugebauer 1991).
\end{enumerate}

Fig. 6 gives the luminosity sequence from Guiderdoni {\it et al.} (1998).
Franceschini {\it et al.} (1991, 1994, and other papers of this series) 
follow the same scheme
for items 1, 2, 3, 4, and use slightly different IR/submm templates, 
whereas Fall {\it et al.} (1996) basically use constant 
dust--to--metal ratios and black--body spectra. Recently, 
Silva {\it et al.} (1998) proposed a more sophisticated treatment 
in which transfer is computed in molecular clouds.
The method of Guiderdoni {\it et al.} (1998) has been extended to 
obtain far--UV 
to radio spectra, and to study local templates by Devriendt {\it et al.} 
(1999). Fig. 7 shows the predicted optical/IR/submm spectrum of an ULIRG, 
and fig. 8 and 9 display examples of fits for observed objects 
(a spiral galaxy and an ULIRG) from this latter paper.

These spectra can subsequently be used to model IR/submm counts.
The simplest idea is to implement luminosity and/or number evolution 
which are parameterized as power laws of $(1+z)$ 
(e.g. Blain and Longair 1993a, Pearson \& Rowan--Robinson 1996 ; see
also references in Lonsdale, 1996). These power laws are generally derived from
fits of the slope of {\sc iras} faint counts (which do not probe deeper 
than $z \simeq0.2$). Then they are extrapolated up to redshifts of a few
units. Unfortunately, various analyses of {\sc iras} deep counts yield 
discrepant 
results at $S_{60} < 300$ mJy, and the amount of evolution is a matter 
of debate (see e.g. Bertin {\it et al.} 1997, for a new analysis and
discussion). This uncertainty increases in the extrapolation 
at higher $z$.

\section{Semi--analytic modelling}

These classes of models assume that all galaxies form at the same 
redshift $z_{for}$. But the paradigm of the hierarchical growth of structures
implies that there is no clear--cut redshift $z_{for}$. In this paradigm,
the modelling of dissipative and non--dissipative processes 
ruling galaxy formation (halo collapse, cooling, star formation,
stellar evolution and stellar feedback to the interstellar medium) has been
achieved at various levels of complexity, in the so--called {\it
semi--analytic} approach which has been successfully applied to the prediction
of the statistical properties of galaxies (White \& Frenk 1991; Lacey \&
Silk 1991, Kauffmann {\it et al.} 1993, 1994; Cole
{\it et al.} 1994; Somerville \& Primack 1999 ; see other papers of these 
series). In spite of 
differences in the details, the conclusions of these models in the UV, visible
and (stellar) NIR are remarkably similar.

A first attempt to compute the IR evolution of galaxies with the 
Press--Schechter formalism has been proposed by Blain \& Longair (1993a,b), 
but with a crude treatment of dissipative processes.
In Guiderdoni {\it et al.} (1996, 1997, 1998), we extend this approach
by implementing spectral energy distributions in the IR/submm range. 
As a reference, we take
the standard CDM case with $H_0$=50 kms$^{-1}$ Mpc$^{-1}$, $\Omega_0=1$, 
$\Lambda=0$ and $\sigma_8=0.67$.
We assume a Star Formation Rate $SFR(t)=M_{gas}/t_*$, with $t_* \equiv \beta
t_{dyn}$. The efficiency parameter $1/\beta =0.01$ gives a nice fit of
local spirals. The robust result of this type of modelling is a
cosmic SFR history that is too flat with respect to the data.

\begin{figure}[htb]
\centerline{
\psfig{figure=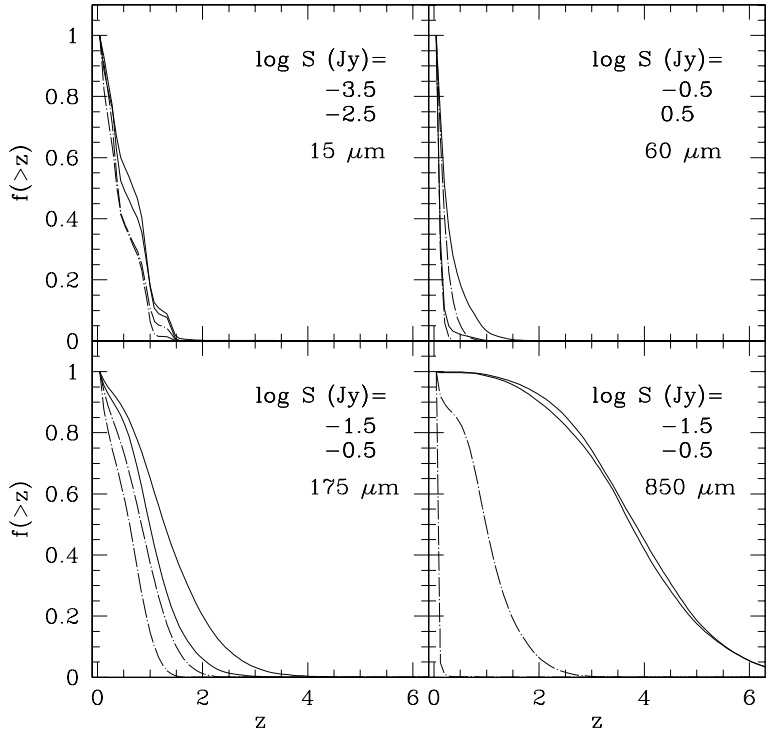,width=0.8\textwidth}}
\caption{\small Redshift distributions at 15 $\mu$m, 60 $\mu$m, 175 $\mu$m, 
and 850 $\mu$m for models A (dashes and dots) and E (solid line). Model E is 
normalized to reproduce the CIRB. The average values of $\log S_\nu$
(in Jy) for $\Delta \log S_\nu = 1$ flux bins are plotted for each curve. 
For instance, $-0.5$ stands for 100 mJy $<S_\nu <$ 1 Jy.}
\end{figure}

As a phenomenological way of
reproducing the steep rise of the cosmic SFR history from $z=0$ to $z=1$, 
we introduce a ``burst'' mode of star formation 
involving a mass fraction that increases with $z$
as $(1+z)^4$ (similar to the pair rate), with ten times higher 
efficiencies $1/\beta=0.1$.
The predicted cosmic SFR density is given in fig. 5.
The resulting model, called ``model A'' does not predict enough flux
at IR/submm wavelengths to reproduce the level of the CIRB. This is sort
of a minimal model that fits the COB and extrapolates the IR/submm fluxes
from the optical (see section 4 on extinction).
In order to increase the IR/submm flux, we have to assume that a small 
fraction of the gas mass (typically less than 10 \%) is involved in
star formation with a top--heavy IMF in heavily--extinguished objects 
(ULIRG--type galaxies). Then a  sequence of models is derived. The amount
of ULIRG can be normalized e.g on the {\it local}
{\sc iras} luminosity function (to reproduce the {\sc iras} bright counts) and
on the level of the CIRB. 
The so--called ``model E'' normalized to the flux level determined in
Guiderdoni {\it et al.} (1997) (see fig. 3) is hereafter 
used to predict faint counts and redshift distributions. An extension 
to NIR and visible counts is given in Devriendt \& Guiderdoni (1999).

\begin{figure}[htb]
\centerline{ 
\psfig{figure=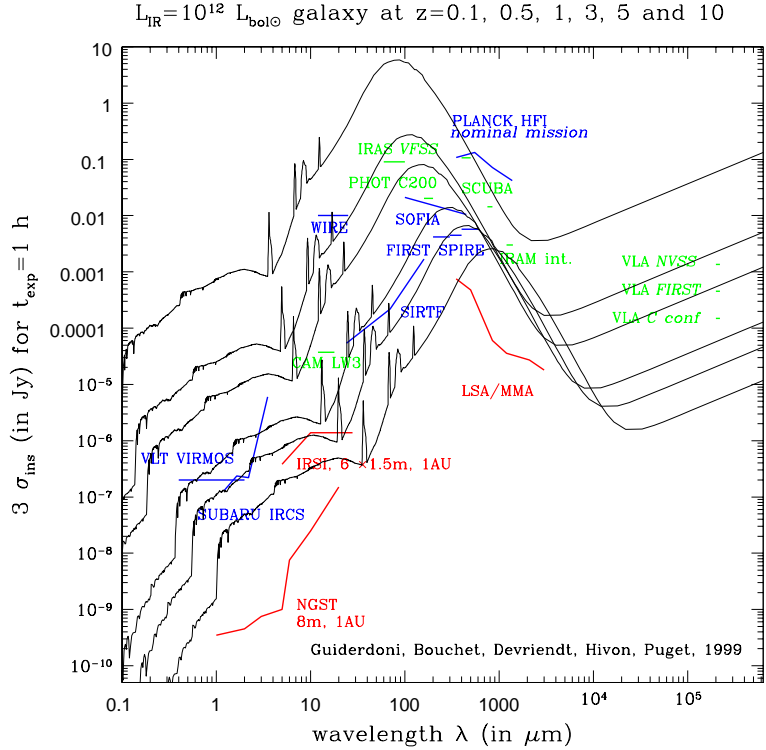,width=0.8\textwidth}}
\caption{\small Observer--frame model spectrum of a $L_{IR}=10^{12} L_{\odot}$
ULIRG at increasing redshifts (from top to bottom), computed by 
Devriendt {\it et al.} (1999). The instrumental 
sensitivities of various past, current, and forthcoming instruments 
(ground--based and satellite--borne telescopes)
are plotted. This figure taken from Guiderdoni {\it et al.} (1999)
is available upon request to {\tt guider@iap.fr}.
Upgrading of the instrumental sensitivities can be achieved when it is 
necessary.}
\end{figure}

Fig. 10 and 11 give the predicted number counts and redshift distributions
at 15, 60, 175, and 850 $\mu$m for this model, that have been produced in
Guiderdoni {\it et al.} (1998) before the publication of the {\it ISOPHOT} 
and {\it SCUBA} faint counts (their fig. 17). 
The agreement of the predicted number counts with the data seems good enough 
to suggest that these counts do probe the evolving 
population contributing to the CIRB. The model
shows that 15 \%  and 60 \% of the CIRB respectively at 175 $\mu$m 
and 850 $\mu$m are built up by objects brighter than
the current limits of {\it ISOPHOT} and {\it SCUBA} deep fields. 
The predicted median redshift of the {\sc iso}--HDF
is $z \sim 0.8$. It increases to $z \sim 1.2$ for the deep {\it ISOPHOT}
surveys, and to $z \ge 2$ for {\it SCUBA}, though the latter value seems to be
very sensitive to the details of evolution.
The model by Toffolatti {\it et al.} (1998) is also shown in fig. 10. It 
gives more bright sources and less faint sources at submm wavelengths. 
\begin{figure}[htb]
\centerline{
\psfig{figure=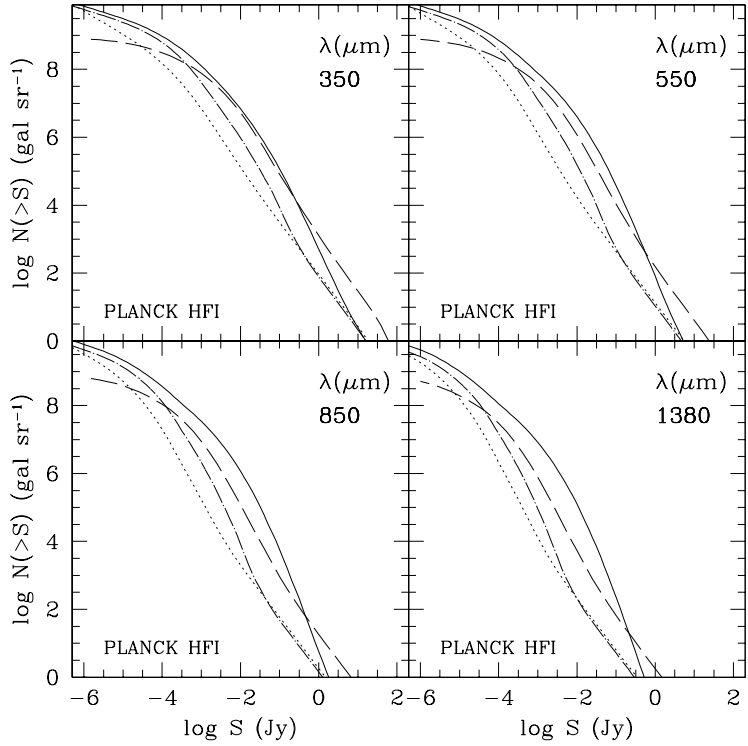,width=0.8\textwidth}}
\caption{\small Predicted counts at {\sc planck} {\it HFI} frequencies
for models A (dashes and dots) and E (solid line). Model E is normalized to 
reproduce the CIRB. Dotted line : no evolution. Dashes : FIR/submm thermal
sources in Toffolatti {\it et al.} (1998).}
\end{figure}

\section{Future instruments}
Fig. 12 gives the far--UV to submm spectral energy distribution that is
typical of a 
$L_{IR}=10^{12}$  $L_\odot$ ULIRG at various redshifts. 
This model spectrum is taken from the computation of 
Devriendt {\it et al.} (1999).
The instrumental sensitivities
of various past and on--going satellite and ground--based
instruments are plotted on this
diagram~: the {\sc iras} {\it Very Faint Source Survey} at 60 $\mu$m,
{\sc iso} with {\it ISOCAM} at 15 $\mu$m, and {\it ISOPHOT} at 175 $\mu$m, 
the IRAM interferometer at 1.3 mm, {\it SCUBA} at 450 and 850 $\mu$m, 
and various surveys with the VLA.
Forthcoming missions and facilities include {\sc wire}, {\sc sirtf}, SOFIA, 
the {\sc planck} {\it High Frequency 
Instrument}, the {\sc first} {\it Spectral and Photometric Imaging REceiver},
and the imaging modes of the SUBARU {\it IRCS} and VLT {\it VIRMOS} 
instruments. Finally, the capabilities of the {\sc ngst}, LSA/MMA, and 
Infrared Space Interferometer ({\sc darwin}) are also plotted.

The final sensitivity of the next--generation instruments
observing at IR and submm wavelengths ({\sc wire}, {\sc sirtf}, SOFIA,
{\sc planck}, {\sc first}) is going to be confusion limited. However,
the observation of a large sample of ULIRG--like objects in the redshift
range 1--5 should be possible. More specifically,
the all--sky shallow
survey of {\sc planck} {\it HFI}, and the medium--deep survey of 
{\sc first} {\it SPIRE}
(to be launched by ESA in 2007), will respectively produce bright
($S_\nu >$ a few 100 mJy) and faint ($S_\nu >$ a few 10 mJy) counts that 
will be complementary. 

\begin{figure}[htb]
\centerline{
\psfig{figure=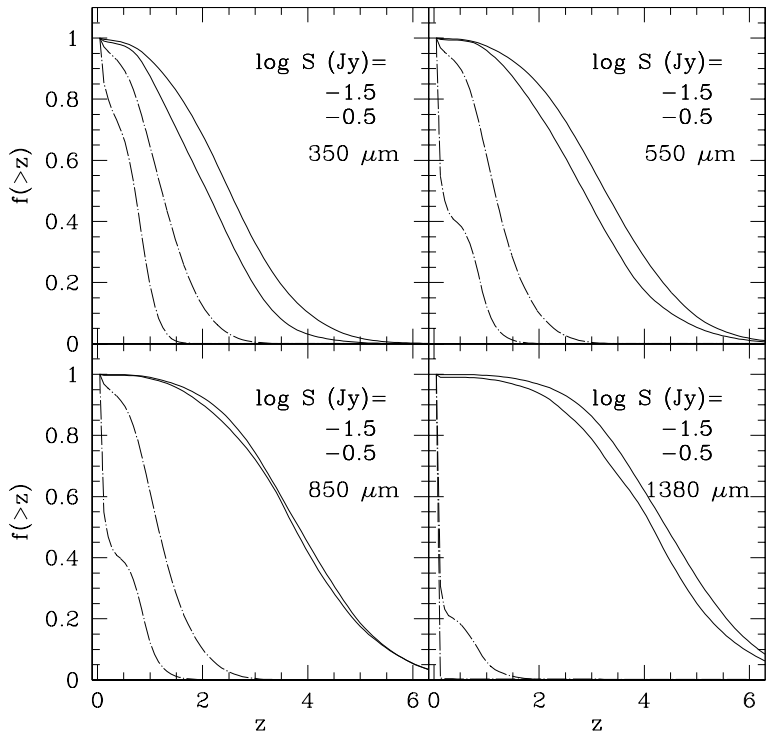,width=0.8\textwidth}}
\caption{\small  Redshift distributions at {\sc planck} {\it HFI} frequencies, 
for models A (dashes and dots) and E (solid line). Model E is 
normalized to reproduce the CIRB. The average values (in Jy) for 
$\Delta \log S_\nu = 1$ flux bins are plotted for each curve.
For instance, $-0.5$ stands for 100 mJy $<S_\nu <$ 1 Jy.}
\end{figure}

Table 1 resumes the various sources of fluctuation in the six bands of 
{\sc planck} {\it HFI} (Bersanelli {\it et al.} 1996, and {\sc planck} {\it HFI}
Consortium, 1998). The confusion limit due to unresolved sources in beam
$\Omega$ has been
roughly estimated with the theoretical faint counts from model E, according
to the formula 
$\sigma_{conf} = (\int_0^{S_{lim}} S^2 (dN/dS) dS \Omega)^{1/2}$.
The values $\sigma_{conf}$ and $S_{lim}=q\sigma_{tot}$ have been computed
iteratively with $q=5$. The 1$\sigma$ total fluctuation is
$\sigma_{tot}=(\sigma_{ins}^2 + \sigma_{conf}^2 + \sigma_{cir}^2
+ \sigma_{CMB}^2)^{1/2}$. However, this does not give a sound estimate of what 
will actually be possible with {\sc planck}, once proper algorithms of 
filtering and source extraction are implemented. Source extraction can be 
studied on simulated maps. Tegmark \& de Oliveira--Costa (1998) showed 
that (i) $\sigma_{tot}$ is about 40 to 100 mJy for the {\it HFI} frequencies 
after filtering, (ii) 40 000 and 5000 sources are expected (at the $5\sigma$
level in 8 sr) respectively at 857 GHz and 545 GHz, and (iii) the CMB 
reconstruction is not jeopardized by the presence of point sources
at the level predicted by the models. Similar results are obtained
by Hobson {\it et al.} (1998b) with a maximum--entropy method 
on mock data (Hobson {\it et al.} 1998a).
In both estimates, source counts 
and source confusion are based on the predicted counts of Toffolatti et al. 
(1998), which differ from ours : Toffolatti's counts are higher at the bright 
end, and fainter at the faint end ; hence his source confusion is lower.
Table 1 gives the number densities expected with flux limits of 
100 mJy and 500 mJy, according to Guiderdoni {\it et al.} (1998).
The reader is invited to note the strong sensitivity of the counts to the 
exact flux limit. So the expectations of source counts with {\sc planck} 
(and {\sc first}) are severely model--dependent. The on--going follow--up of 
{\it ISOPHOT} and {\it SCUBA} sources will eventually give redshift
distributions that will strongly constrain the models and help improving 
the accuracy of the predictions.

As far as {\sc first} is concerned, a 10 deg$^2$ survey with {\it SPIRE} will
result in $\sim 10^4$ sources ({\sc first} {\it SPIRE} Consortium, 1998).
The study of the $250/350$ and $350/500$ colors are suited to point out 
sources which are likely to be at high redshifts. These sources can be 
eventually followed at 100 and 170 $\mu$m by the {\sc first} 
{\it Photoconductor Array Camera \& Spectrometer} and by the FTS mode
of {\it SPIRE}, to get the spectral energy distribution at $200 \leq \lambda 
\leq 600$ $\mu$m with a typical resolution $R\equiv \lambda /\Delta\lambda=20$.
After a photometric and spectroscopic follow--up, the submm observations will
readily probe the bulk of (rest--frame IR) luminosity associated with
star formation. The reconstruction of the cosmic SFR comoving density will 
thus take into account the correct luminosity budget of high--redshift 
galaxies.
However, the spatial resolution of the submm instruments will be limited, 
and only the LSA/MMA
should be able to resolve the IR/submm sources and study the details of their 
structure.

\begin{table}[htb]
\caption[]{Sources of fluctuations in {\sc planck} {\it HFI} wavebands.
(1) Central frequency in GHz. (2) Beam full width half maximum in arcmin.
(3) $1\sigma$ instrumental noise for 14 month nominal mission.
(4) $1\sigma$ fluctuations due to cirrus at $N_{HI} = 1.3~10^{20}$ cm$^{-2}$
(level of the cleanest 10 \% of the sky). The fluctuations
have been estimated following Gautier {\it et al.} (1992) with $P(k) \propto
k^{-2.9}$ and $P_{0, 100 \mu m} = 1.4~10^{-12} B_{0, 100 \mu m}^3$.
(5) $1\sigma$ CMB fluctuations for $\Delta T /T =10^{-5}$.
(6) $1\sigma$ confusion limit due to IR sources in beam $\Omega \equiv
\theta_{FWHM}^2$.
(7) $\sigma_{tot}=(\sigma_{ins}^2 + \sigma_{conf}^2 + \sigma_{cir}^2
    + \sigma_{CMB}^2)^{1/2}$. Here $\sigma_{cir}$ is for 
    $N_{HI} = 1.3~10^{20}$ cm$^{-2}$.
(8) Number counts at a flux limit of 100 mJy in 1 sr.
(9) Number counts at a flux limit of 500 mJy in 1 sr.}
\begin{center} \scriptsize
\begin{tabular}{rrrrrrrrr}

\hline

$\nu$ & $\theta_{FWHM}$ & $\sigma_{ins}$ & $\sigma_{cir}$ & $\sigma_{CMB}$ &
$\sigma_{conf}$ & $\sigma_{tot}$ & $N(>{\rm 100 mJy})$ & $N(>{\rm 500 mJy})$\\

GHz & arcmin & mJy & mJy & mJy & mJy & mJy & in 1 sr & in 1 sr\\ 
(1) & (2) & (3) & (4) & (5) & (6) & (7)  & (8)  & (9)\\

\hline

857 & 5 & 43.3    & 64     & 0.1 & 146 & 165  & 120000 & 2500 \\
545 & 5 & 43.8    & 22     & 3.4 & 93  & 105  &  37000 & 1200 \\
353 & 5 & 19.4    &  5.7   & 17  & 45  & 53   &   6800 & 20   \\
217 & 5.5 & 11.5  &  1.7   & 34  & 17  & 40   &    250 & 1    \\
143 & 8.0 & 8.3   &  1.4   & 57  & 9.2 & 58   & --     & -- \\
100 & 10.7 & 8.3  &  0.8   & 63  & 3.8 & 64   & --     & -- \\

\hline 
\end{tabular}
\end{center}

\end{table}

\section{Conclusions}
\begin{enumerate}
\item High--redshift galaxies emit much more IR than predictions 
based on the local IR luminosity function, without evolution. 
The submm counts start unveiling the bright end of the population that is 
responsible for the CIRB. The issue of the relative contributions of the 
starbursts and AGNs to dust heating is still unsolved. Local ULIRGs
(but the brightest ones) seem to 
be dominated by starburst heating. However the trend at higher redshift is 
unknown. 
\item It is difficult to correct for
the influence of dust on the basis of the optical spectra alone.
Multi--wavelength studies are clearly necessary to address the 
history of the cosmic SFR density. Forthcoming instrument will help
us greatly improve our knowledge of the optically dark side of galaxy
formation. Next milestones are {\sc sirtf}, SOFIA, the {\sc planck} 
{\it High Frequency Instrument}, the {\sc first} {\it Spectral and 
Photometric Imaging REceiver}, and the LSA/MMA.
\item Under the assumption that starburst heating is dominant, simple models 
in the paradigm of hierarchical clustering do reproduce the current 
IR/submm data. These models normalized by means of the current and 
forthcoming counts should help us predict the number of IR/submm sources
that will be observed by the {\sc planck} {\it High Frequency Instrument}, the 
contribution of the unresolved sources to the submm anisotropies, and the 
final strategy for foreground separation and interpretation. Though source 
counts are strongly model--dependent, and only partly constrained by the 
current set of data, the studies so far seem to show that the quality of the 
reconstruction of CMB anisotropies is not severely degraded by the presence 
of foreground point sources.
\end{enumerate}

\acknowledgments

I am grateful to F.R. Bouchet, J.E.G. Devriendt, E. Hivon, 
G. Lagache, B. Maffei, and J.L. Puget who collaborated to many aspects of 
this program, as well as to H. Dole and the {\sc FIRBACK} consortium 
for illuminating discussions. My thanks also to
G. De Zotti, A. Franceschini, and L. Toffolatti.

\end{document}